# Effect of Range Naming Conventions on Reliability and Development Time for Simple Spreadsheet Formulas

Ruth McKeever, Kevin McDaid

Dundalk Institute of Technology,

Dundalk, Ireland.

catherineruth.mckeever@dkit.ie, kevin.mcdaid@dkit.ie

**ABSTRACT**

*Practitioners often argue that range names make spreadsheets easier to understand and use, akin to the role of good variable names in traditional programming languages, yet there is no supporting scientific evidence. The authors previously published experiments that disproved this theory in relation to debugging, and now turn their focus to development. This paper presents the results of two iterations of a new experiment, which measure the effect of range names on the correctness of, and the time it takes to develop, simple summation formulas. Our findings, supported by statistically significant results, show that formulas developed by non-experts using range names are more likely to contain errors and take longer to develop.*

*Taking these findings with the findings from previous experiments, we conclude that range names do not improve the quality of spreadsheets developed by novice and intermediate users.*

*This paper is important in that it finds that the choice of naming convention can have a significant impact on novice and intermediate users' performance in formula development, with less structured naming conventions resulting in poorer performance by users.*

## 1      INTRODUCTION

With the current estimate of Excel users worldwide estimated at 400m the importance of spreadsheets cannot be overstated. In 2007, Microsoft sold 71 million licenses for Microsoft Office [Microsoft, 2007]; in 2008 they sold 120 million [Microsoft, 2008]. An investigation into the importance and use of spreadsheets in the City of London [Croll, 2005] is summed by the following quote: "Excel is utterly pervasive. Nothing large (good or bad) happens without it passing at some time through Excel". Spreadsheet development is increasingly recognized by authors as programming. For example, Burnett et al [2004] declare that "spreadsheet languages are the most widely used end-user programming languages to date—in fact, they may be the most widely used of *all* programming languages". Panko and Halverson [1996] state that "quite simply, spreadsheeting is quite a bit like programming".

The power and flexibility of spreadsheets make them an indispensible tool in modern business, at the cost of reliability. The ease with which any user can become a developer leads to widespread uncontrolled use, as they are not subject to rigorous testing procedures favoured by professional programmers. Consequently, spreadsheet development is known to be highly unreliable. Spreadsheets that are unregulated and untested are used for critical decision-making, leading to media reports of accidental and intentional errors leading to fiscal and reputational losses. A number of examples can be found on the EuSpRIG website [EuSpRIG, 2010]. Reliability is arguably the most important aspect of spreadsheet quality. The reliability of a spreadsheet is measured by the accuracy of the information produced, and is damaged by the number and magnitude of errors that commonly occur. Range names are frequently recommended by academics and practitioners to improve the reliability of spreadsheets, yet no empirical evidence has been cited to support these recommendations.

The purpose of this study is to evaluate empirically the effect of different range naming conventions on spreadsheet reliability, and development time, by conducting two experiments. Both experiments have the same design, however the second improves on certain issues that arose during the first. This work represents the first scientific attempt to establish the benefits or otherwise of range names in the development of a spreadsheet. It is important in the context of the extensive use of spreadsheets and the recommendations of practitioners. Furthermore, when combined with earlier work, the paper presents important conclusions on the use of range names by non-expert users. The findings are clear; contrary to published opinion we find no evidence to endorse the use of range names in the development of reliable spreadsheets by novice and intermediate users. In particular, we find that poorly structured names have a particularly harmful effect on formula quality. While this may seem obvious to a practitioner, this paper presents empirical evidence to support this finding.

Section 1 of this paper begins by summarizing the important literature in this area, including experiments previously published by the authors that investigated the impact of range names on debugging. Section 3 details the methodology behind the experiment design, and in Section 4 the authors present in detail the results of both experiments. Section 5 comprises a discussion of the results, and Section 6 contains the conclusion, including proposals for future work.

## 2    LITERATURE REVIEW

It is widely acknowledged that spreadsheet use is ubiquitous in industry, however quality is often overlooked. Researchers frequently suggest that software programming practices should be applied to spreadsheets, in an attempt to advance quality. For example, "if our goal is to teach spreadsheeters how to develop spreadsheets professionally, we may be able to draw on what we already know from program development" [Panko and Halverson, 1996]. One such suggestion is that range names could be used in place of cell references, transferring the advantages of good variable names to the spreadsheet domain.

**Naming of Cell Ranges**

Range names are a powerful feature of Excel. In a spreadsheet, a range is defined as a cell or group of cells. A range name is a name given to a range, which can then be used throughout the workbook in place of the cell reference. It is often suggested that they can make a spreadsheet easier to understand and to develop. There is no existing research to back up this suggestion, and the aim of this project is to carry out the research to test this theory.

The advantages of naming conventions in programming are the subject of many studies. Keller [1990] established that people have less difficulty reading programs that follow a defined naming scheme, although they cannot explain why they find them easier to read. The legacy of using one-letter variables in mathematics is thought to be one cause of poor naming [Rowe, 1985]. An interesting theory is spelling mistakes: "if people make spelling mistakes for words whose correct spelling they have seen countless times, it is certain that developers will make mistakes, based on the same reasons, when typing a character sequence they believe to be the spelling of an identifier" Jones [2008].

Range names are recommended by a number of authors, in published conference papers and journals. For example, Bewig [2003] states: "although it takes a little more work initially to create names, it should be clear they make formulas easier to write and to read. This is especially true in large spreadsheets where you may have scores of references." The same author, in a subsequent paper [Bewig, 2005], advocates the proper construction of range names for eliminating the problem of referring to the wrong cell while constructing formulas, and states "well chosen names are the first and best form of documentation". Grossman et al. [2009] illustrate how range names can be used to replace complex nested-if formulas with the lookup technique. As nested-if formulas are considered risky, the lookup technique uses names to make the logic required simpler, more visible and therefore less risky.

Range names are also frequently discussed by practitioners, in newsletters, websites and blogs. In CompAct, a newsletter for the Society of Actuaries, Campbell [2009] makes the extreme statement, that in order to improve adaptability "any cells being used in formulas should be referred to as named ranges." An example of expert practitioners who have found names to be particularly useful is OPERIS. A spreadsheet consultancy firm based in London, OPERIS have made range names central to their spreadsheet development methodology. Support is provided by OAK, the OPERIS Analysis Kit [OPERIS, 2009] which has many features for handling names.

Not all practitioners are in favour of names. Panko and Ordway [2005] warn that range names "should be considered potentially dangerous until research on using range names is done." Blood [2006] criticizes names for making formulas needlessly long and difficult to audit, stating that names are unnecessary if spreadsheet models are well designed.

**Range names and debugging**

Previously the authors carried out two experiments [McKeever and McDaid, 2010, McKeever et al., 2009] that investigated the impact of range names on the debugging performance of novice users. These were both adapted from a design first used in a study by Howe and Simkin [2006], and later used by Bishop and McDaid [2007]. Participants were given a spreadsheet seeded with errors, and were asked to correct any mistakes they could find, directly in the spreadsheet. They were not told how many errors were in the spreadsheet, or what types of errors were included. Their cell click times were recorded by T-CAT, a "time-stamped cell activity tracking tool" [Bishop and McDaid, 2008].

The results of both these experiments reveal that range names are a hindrance to novices either debugging or developing a spreadsheet. There was a reduction in overall error finding performance in both debugging trials, indicating that the problem is not just in how names are used in a formula, but with their overall inclusion in the spreadsheet. The failure to improve reliability or speed in the development task indicates that it is dangerous to recommend them to novice or intermediate users.

**Summary**

Extensive research has been carried out on spreadsheet errors, and concluded that spreadsheets have poor levels of reliability. Range names are frequently mentioned in the literature as a potential risk mitigation measure. Most of the experts, with a few notable exceptions, recommend their use, however there is no research at present that examines in a structured and methodical manner the impact of names on spreadsheet quality.

The authors previously showed that range names do not improve the debugging performance of novice spreadsheet users. Debugging, however, is only one aspect of spreadsheet programming, arguably less important than development. As formulas have the greatest potential for material error in a spreadsheet, we designed an experiment that directly compared the reliability of formulas developed with range names to formulas developed with cell references.

The experiment described in this paper is part of a wider research project that investigates the impact of range names on different aspects of spreadsheet quality. The authors began by focusing on spreadsheet reliability, first debugging and now formula development. We hope to continue this research and evaluate the effect of range names on all aspects of spreadsheet development and use. Future research plans are described in more detail in Section 6.1.

## 3 METHODOLOGY

The aim of this experiment is to evaluate the reliability and completion time of simple formulas, and how this was influenced by using different types of naming structure. We began by investigating the null hypothesis, and from this developed a number of specific research questions, set out below.

### 3.1 Hypothesis and Research Questions

The following hypothesis was chosen as the basis for this study:

> *Range names, regardless of their structure, have no impact on the reliability and completion time of simple spreadsheet formulas, developed by novice and intermediate users.*

From this hypothesis we derived the following research questions:

*RQ1:* Do users make more mistakes using range names or cell references, when asked to develop a simple spreadsheet formula?

*RQ2:* Does the time it takes users to develop a simple spreadsheet formula differ for formulas using range names than for formulas using cell references?

Range names, as with programming variables, can be chosen according to various conventions. The work will examine each of the research questions above for each of the following six range naming structures:

   a) *Where no two names begin with the same word.*
   b) *Where several different names begin with the same word, but end in a different word.*
   c) *Where several different names begin and end in the same words, with a change in the number in the middle of the name.*
   d) *Where names begin with the same word with a change in the trailing number.*
   e) *Where several different names begin and end with the same word, with a change in the word in the middle.*
   f) *Where names do not follow any naming convention, and are inconsistent.*

The measurement for RQ1 is the number of correct formulas developed. The measurement for RQ2 is the time it takes to develop each formula.

### 3.2 Task Design

It was decided that the most appropriate way to address these questions would be to isolate a basic formula task, using names that followed the naming conventions specified above. Simple addition through use of the additive operator was chosen, as it is one of the most basic, well-known actions in Excel. This does not require any knowledge of Excel functions, and no users implemented the formula using the inbuilt SUM function. The spreadsheet was designed so that the participants would not require domain knowledge in order to understand the spreadsheet and the tasks.

One spreadsheet was used in this experiment, containing six worksheets. Each worksheet contained two identical tasks, based on different lists of data. For one list the participants were required to use range names for the task, for the other, cell references. Each sheet addresses one of the six naming structures, in the same order as they are listed above. All these structures were based on a combination of row and column headings. This resulted in a total of 12 tasks, six with range names, and six with cell references for each participant. Each task requires the user to select seven cells or range names.

*Sheet 1:* Very simple names, that follow a strict naming convention. No two names begin with the same word. Examples include *ArnottsSales, ClearysSales, TopshopSales*.

*Sheet 2:* Two names can begin with the same word. Examples include *TopshopGP, TopshopNP*. After the first implementation of this experiment, it was decided that *GP* and *NP* were too similar, as there was only a difference of one letter and the last letter of each name was the same. The words used were changed to Gross and Net, for example, *TopshopGross* and *TopshopNet*.

*Sheet 3:* Names follow a slightly more complex naming convention, whereby two names can both begin and end in the same word, but contain a different number in the middle. Examples include *HMV2008Profits, HMV2009Profits*.

*Sheet 4:* Names begin with the same word, but end in a different number. Examples include *PrimarkTax2006, PrimarkTax2007*.

*Sheet 5:* Names begin and end in the same words, but have a different word in the middle. Examples include *GAPSecWages, GAPSupWages*. After the first implementation of this experiment, it was decided that *Sup* and *Sec* were also too similar, hence these words were changed to *Assist* and *Clerk*.

*Sheet 6:* Names that do not follow any naming convention. Examples include *CostsQuicksilverFixed, QuicksilverVariableCosts*.

It was decided that sheet level names were most suitable, as the user would only be able to see the names that were developed for the particular task on which they are working. This was to avoid confusion, as there were 264 names in the spreadsheet. The aim was not to confuse the users, but to examine the naming conventions in isolation.

As described earlier, the study was structured on the basis of a within subject design where each participant implemented a simple formula using both cell reference and range names for each of the six range name types identified in the research questions. The order in which the users carried out the tasks was varied by dividing the subjects randomly into two groups i.e. Group A began with range names, then used cell references; Group B began with cell references then moved on to range names. This was done to factor out any learning effect, i.e. that the users might be better prepared to carry out the second task on each sheet, as they had learned what to do from the first task.

### 3.3 Operational Context

This experiment was carried out in a computing laboratory in Dundalk Institute of Technology. Each participant used the PCs installed in this room, running Windows 7 and Microsoft Excel 2007. The participants were randomly assigned to groups A and B, and the researcher explained how to do the experiment. They were then observed whilst carrying out the tasks. As in previous experiments, the T-CAT macro ran in the background for all the participants, and recorded the time it took to complete each task.

The participants in the first implementation of this experiment, Group 1, were 15 postgraduate students from the Higher Diploma in Computing class in Dundalk Institute of Technology, Ireland. Most of them had returned to education after a period in the workplace. The participants in the second implementation, Group 2, were 17 second-year Software Development students, also from Dundalk Institute of Technology. Based on a background survey the participants in both groups were judged to be a mix of novice and intermediate users.

### 3.4 Example Tasks

Shown in Table 1 are two examples of tasks that were given to the participants in Group A. These tasks were included on Sheet 5. The participants in Group B were given the same two tasks, except in Task 3 they were asked to use cell references, and in Task 4, range names. Examples of incorrect formulas developed as answers to these tasks are displayed in the following section.

| **Task 3:** With a formula that uses range names, in cell C28 calculate the total of the following store profits:<br>    Net profit for Powerhouse<br>    Gross profit for Pumpkin Patch<br>    Gross profit for Austin Reed<br>    Net profit for Zara<br>    Net profit for Quicksilver<br>    Gross profit for Quicksilver<br>    Net profit for Topshop | **Task 4:** With a formula that uses cell references, in cell C57 calculate the total of the following store profits:<br>    Net profit for Dillons<br>    Net profit for Dixons<br>    Gross profit for Dillons<br>    Net profit for Liberty<br>    Gross profit for The Paper Mill<br>    Gross profit for Boots<br>    Net profit for Etam |

**Table 1 - Example Tasks**

### 4 RESULTS AND DISCUSSION

This section first examines the amount of errors made in each task, looking at whether more incorrect formulas were developed using range names or cell references. Next, the different type of errors that occurred are examined, with a focus on whether certain errors were more likely where names were used. Finally there is an analysis of the time it took the subjects to develop the formulas for each task.

**Errors made per task**

Table 2 shows the number of subjects who created incorrect formulas in each of the sheets on the spreadsheet, according to whether the task included range names or cell references.

| *Errors* | *Group 1 (15 subjects)* | | *Group 2 (17 subjects)* | |
| --- | --- | --- | --- | --- |
| *Sheet* | *Named Ranges* | *Cell References* | *Named Ranges* | *Cell References* |
| Sheet 1 | 0 | 1 | 0 | 1 |
| Sheet 2 | 4 | 0 | 2 | 2 |
| Sheet 3 | 3 | 0 | 1 | 1 |
| Sheet 4 | 0 | 1 | 0 | 1 |
| Sheet 5 | 3 | 2 | 3 | 0 |
| Sheet 6 | 2 | 0 | 4 | 1 |
| **Total** | **12** | **4** | **10** | **6** |

**Table 2 - Errors per Task**

The results indicate that participants were less effective at developing formulas using range names than using cell references. For the first group statistical analysis, using McNemar's test for paired proportions indicates that there is significant evidence that in the case of Sheet 2 and Sheet 3 users were more effective at developing simple formulas using cell references as opposed to range names. The statistical tests for Group 1 support the following statement, referring to RQ1:

*Novice and intermediate users make fewer mistakes when developing formulae using cell references than using range names where:*

   a) *Several different names begin with the same word, but end in a different word.*
   b) *Several different names begin and end in the same words, with a change in the number in the middle of the name.*

For Group 2, there is statistically significant evidence that in the case of Sheet 5 users were more effective at using cell references to develop simple formulas. The statistical tests for Group 2 support the following statement, referring to RQ1:

*Novice and intermediate users make fewer mistakes when developing formulae using cell references than using range names where several different names begin and end with the same word, with a change in the word in the middle.*

The fact that these findings were inconsistent between the two experiments indicates that the changes made to the experiment design had an impact on the users' performance. In Sheet 5 the words 'Sup' (supervisor) and 'Sec' (secretary) were changed to 'Assist' (assistant) and Clerk, as 'Sup' and 'Sec' were deemed too similar to represent a change in the middle word of the range name. It appears that the similarities in these words also served to confuse the subjects when using cell references, and the change to more obviously different words highlighted the effect of the naming convention, as opposed to the effect of the actual word. In Sheet 2 the naming convention was changed for the second experiment to replace the terms 'NP' (net profit) and 'GP' (gross profit) with 'Net' and 'Gross' respectively, as 'NP' and 'GP' were not deemed sufficiently different to represent a change in the final word in the naming structure. This may explain why the subjects in Group 2 made fewer errors than the subjects in Group 1, however the column headings did not change, as they did in Sheet 5, hence we cannot offer an explanation for the increase in cell reference based errors. Likewise, no changes were made to Sheet 3.

For two naming conventions no mistakes were made using range names, while one mistake was made using cell references for each of Groups 1 and 2, although this is not statistically significant. From Group 1, for two naming conventions there were statistically significant results to conclude that they negatively affect the reliability of the formula; from Group 2, there was statistically significant results to conclude that one naming convention negatively affects the reliability of the formula. These conflicting results suggest that naming conventions have an impact on the effectiveness of range names. Taking the results of the six naming conventions combined there is evidence to suggest that range names should not be used in formulas.

**Error Types**

While the results detailed above refer to the number of incorrect formulas developed by the subjects, it is worth looking at the actual formulas to see what type of errors were made. In several cases the offending formula contained more than one error.

It was expected that the participants would make selection and omission errors under both conditions. Selection errors occur when the wrong cell or name is used in a formula; omission errors occur when a reference is left out. Looking at both Groups 1 and 2 together, of the 22 incorrect formulas developed using range names 10 contained selection errors and 14 contained omission errors. Of the 9 incorrect formulas developed using cell references, 2 contained selection errors, and 7 contained omission errors.

An unexpected finding was that some participants, while working with range names, made another type of error: in three cases a subject added an extra name to the formula. During Task 3 a subject from Group 1 added the name ZaraNP twice, and during Task 6 a subject from Group 1 included Costcutter2008Profits as well as Costcutter2009Profits. During Task 9, a subject from Group 2 included OasisClerkWages as well as OasisAssistWages. This type of error did not occur during the cell reference tasks.

Below are two examples of formulas that contain omission errors developed as answers by participants from Group 1 to the two tasks detailed in the previous section:

One participant, from Group 1, gave the following erroneous answer to Task 3:

=PowerhouseNP+PumpkinPatchGP+AustinReedGP+ZaraNP+ZaraNP+QuicksilverNP+TopshopNP

The correct answer would also include "QuicksilverGP", and omit one instance of "ZaraNP".

Another participant, from Group 2, gave the following erroneous answer to Task 4:

=SUM(B35,B36,B41,C48,C51,B38)

The correct answer should include cell C35, the value for Dillons Gross profit.

**Time Results**

Table 3 shows the average time in minutes it took the participants to complete the tasks on each sheet, according to whether the task included range names or cell references. It took Group 1 on average 0.7 minutes longer, and Group 2 0.77 minutes longer, to complete the tasks when range names were used. These times include the assimilation time for both beginning the experiment, and for each sheet. It can be presumed that the subjects would take less time to complete the second task on each sheet, as they are already familiar with the sheet since doing the first task. Dividing the participants randomly into groups A and B so that some would complete the name tasks first and others would complete the cell reference tasks first eliminated this bias.

| *Times* | *Group 1* | | *Group 2* | |
|---|---|---|---|---|
| *Sheet* | *Named Ranges* | *Cell References* | *Named Ranges* | *Cell References* |
| Sheet 1 | 1.98 | 1.33 | 1.74 | 1.74 |
| Sheet 2 | 1.95 | 1.30 | 2.43 | 1.56 |
| Sheet 3 | 1.45 | 1.10 | 1.65 | 1.50 |
| Sheet 4 | 1.35 | 1.04 | 1.57 | 1.05 |
| Sheet 5 | 1.31 | 1.19 | 1.38 | 1.20 |
| Sheet 6 | 3.19 | 1.04 | 4.00 | 1.10 |
| **Total** | **1.87** | **1.17** | **2.13** | **1.36** |

**Table 3 - Average Time (Minutes)**

Statistical analysis of the significance of the difference in the times it took to perform each task was performed using a paired T- test. For Group 1 these indicate that the time it took users to complete the tasks associated with sheets 2, 3, 4 and 6 for range names as opposed to cell references was significantly higher at a 5% level of significance. The statistical tests for Group 1 support the following statement, referring to RQ2:

*Novice and intermediate users take longer to develop formulae using range names than using cell references where:*

a) *Several different names begin with the same word, but end in a different word.*
b) *Several different names begin and end in the same words, with a change in the number in the middle of the name.*
c) *Names begin with the same word with a change in the trailing number.*
d) *Names do not follow any naming convention, and are inconsistent.*

For Group 2, the statistical tests indicate that the time it took users to complete the tasks associated with sheets 2, 4, 5 and 6 for range names as opposed to cell references was significantly higher at a 5% level of significance. The statistical tests for Group 2 support the following statement, referring to RQ2:

*Novice and intermediate users take longer to develop formulae using range names than using cell references where:*

a) *Several different names begin with the same word, but end in a different word.*
b) *Names begin with the same word with a change in the trailing number.*
c) *Several different names begin and end with the same word, with a change in the word in the middle.*
d) *Names do not follow any naming convention, and are inconsistent.*

The clear consistency in this result is that for the most structured naming convention there is no significant result to differentiate between development time for range names and cell references. For the least structured names, there are significant results from both experiments that find it takes more time to develop a formula using range names. For the various degrees of structure in-between, the findings are less conclusive, yet indicate that it is faster to develop using cell references.

For each naming convention, the average time to complete the task was higher where the participant used range names. This was statistically significant for four naming structures from each experiment. The number was particularly high for the last structure – where no naming convention was followed. This is not surprising, as the participants did not know what letter each name began with, and hence did not know where to look for the name in an alphabetically ordered list, or what letter to begin typing. This highlights the importance of following some kind of convention, if range names are chosen as a suitable practice. Further investigation should be carried out on the role of alphabetisation and the ease with which subjects can select a name from a list.

One possible reason for the increase in development time using range names is that it would naturally take longer to type a sequence of characters than to click on a cell. All participants were shown how to chose a name by clicking the *Use in Formula* button to display a list of all the names available in the worksheet. We argue that the time it takes to do this is comparable to the time it takes to click a cell. The list of names is sorted alphabetically, making it easier for the user to find the name they are looking for, unlike the list of data displayed on the worksheet, which is not sorted. During the experiment all subjects were observed following this method for choosing a name, instead of typing the name.

## 5    THREATS TO VALIDITY

In this section the authors first discuss the threats to construct validity, followed by internal and external validity.

**Construct Validity**

The authors contend that the hypothesis can be evaluated sufficiently by measuring the number of correct formulas, and the time it takes to develop each formula. One potential issue is that the time measurement is calculated by recording the time spent between the user entering the cell and leaving it after completing the formula. This includes the time it takes the user to read the task and comprehend the model.

Another issue regarding construct validity is the decision to measure performance on a summation task only. A wider selection of tasks would increase construct validity, but reduce internal validity when drawing comparisons between results on such a small sample size.

**Internal Validity**

In the first implementation of this experiment the words chosen for the range names on two sheets were identified as being too similar. While this may be a common occurrence in real world spreadsheets, it affects the internal validity of this study as any error made could be attributed to the similarities rather than the naming convention. Different words were chosen for the second implementation.

One issue with this experiment was that when the user clicked on a cell that was named while writing a formula, the name of the cell appeared in the formula rather than the cell reference. Unfortunately this is not something that can be caught by the T-CAT macro, as cell selections cannot be recorded while a formula is being edited. The participants were given clear instructions, and were closely monitored to ensure that they followed the instructions exactly. For Group 2, the cells that were named were also locked, so that the participants could not click on the cell to choose the reference. One participant tried to type the cell reference manually into the formula, but was immediately asked to use the range names instead. A better solution to this issue must be developed before this experiment could be rolled out to a larger group, as it is unnatural that a user would not be able to click on a cell in a working environment.

**External Validity**

Students were used in these experiments. Although this approach is controversial, we argue that the results can be generalized to the majority of spreadsheet users, but not experts who regularly use range names as part of their development methodology. Studies have shown that students have similar abilities to professionals, for example Galletta et al. [1993] found that spreadsheet experts did not outperform novices in finding spreadsheet formula errors. Considerable research has recognized that spreadsheets are rarely developed by professional programmers. For example, Purser and Chadwick [2006] found that 85% of survey participants developed the spreadsheets that they use. This is not to say that the developers are not professionals, but that their expertise lies within their domain rather than in programming.

The subjects who took part in these trials were taught by the researchers how to use range names. This was necessary for the purposes of the trial, but does not reflect how novice and intermediate practitioners would learn about naming in the real world. As mentioned in Section 5, they were taught how to use the *Use in Formula* feature to choose a name. This may not reflect how names are used in practice.

## 6  CONCLUSION

Described in this paper is a new experiment design, that looks at the reliability of formulas developed using structured and unstructured range names, and the time to develop such formulas. The results of two iterations of this experiment are detailed, and the findings discussed. The authors have three main conclusions.

First, there is no evidence to support the theory that range names lead to more reliable formulas. The subjects in the first group developed twelve incorrect formulas using range names, and four incorrect formulas using cell references. The subjects in the second group developed ten incorrect formulas using range names, and six incorrect formulas with cell references. With range name use there was an increase in the number of omission errors, and subjects made some addition errors, which did not occur with cell references. This leads us to conclude that range names have a negative impact on the reliability of cell formulas developed by novice and intermediate users

Second, we found evidence that formulas developed using range names took longer to develop than formulas using cell references. For the first group, the average time it took do develop each formula was longer for all the range name tasks than the corresponding cell reference tasks. This was statistically significant for four out of six cases. For the second group, the average time it took to develop five out of six range name formulas was longer than the corresponding cell reference formula; the average times were identical for one formula. Again, this is statistically significant in four out of six cases. This leads us to conclude that range names make formulas longer to develop.

Third, we found evidence that the degree to which the range name is structured has an impact on both the reliability of, and the time it takes to develop, the formula. The most structured naming convention did not cause a significant increase in errors, or in time to develop. The formulas developed using names that followed no naming convention took significantly longer to develop, and averaged more errors (although the error rate was not statistically significant). Although the authors do not endorse range name use on the basis of the previous conclusions, if a practitioner decides to use range names in a spreadsheet we recommend they follow a strict naming convention.

The naming conventions that were found to have a significant negative impact on reliability also took significantly longer to use in a formula. This is a clear indictment of their suitability as naming conventions. Importantly, the increase in selection errors illustrates that range names do not help the user to avoid referring to the wrong cell, as is often claimed. The increase in the time it takes to develop a formula dispels the theory that range names make formulas easier to create.

These findings are consistent with the findings of previous experiments, and indicate that range names do not make formulas easier for novice and intermediate users to either develop or debug. The resulting spreadsheets are less reliable than those developed using cell references. This study is part of on-going research into the role of range names in the entire spreadsheet development process. The following section details how the authors plan to proceed with this research.

### 6.1 Future work

Firstly, it is crucial to the validity of this research that the scope of the project be increased to focus on professionals. We also plan to examine how range names are used in practice, and why some practitioners choose to use them. The authors have not yet looked at whether range names reduce development time when the range is located on a different sheet. If this was found to be the case then this time saving might be worthwhile to the developer. If range names do not improve the reliability of spreadsheets, or make them easier to debug, they could be used to improve other aspects of spreadsheet quality, such as understandability.

An alternative route of enquiry would be to investigate why range names so clearly confuse users. After the initial debugging experiments two factors were briefly examined as possible causes. High cognitive load was investigated, as participants must remember what each name refers to when analysing a formula, whereas with a cell reference they can see exactly where is being referred to. Overconfidence was also suggested as a contributing factor. Panko [2003] suggests that overconfidence has an impact on spreadsheet error rates, emphasizing the self reinforcing nature of risky behaviour in relation to spreadsheets: "developers who do not do comprehensive error checking are rewarded both by finishing faster and by avoiding onerous testing work".